\documentclass[twocolumn,twoside,slac_two]{revtex4}
\usepackage{graphicx}
\usepackage{fancyhdr}
\begin{document}

\title{Indirect Search for Dark Matter from the center of the Milky Way with the   Fermi-Large Area Telescope}

%

\author{Vincenzo Vitale}
\affiliation{Istituto Nazionale di Fisica Nucleare, Sez. Roma Tor Vergata, Roma, Italy}
\author{Aldo Morselli, for the Fermi/LAT Collaboration}
\affiliation{Istituto Nazionale di Fisica Nucleare, Sez. Roma Tor Vergata, Roma, Italy}

\begin{abstract}

The Galactic Center region is expected to host the largest density of Dark Matter (DM) particles within the Milky Way.
Then a relatively large gamma-ray signal would be expected from the possible DM particles annihilation (or decay).
We are searching for the DM gamma-ray signal from the Galactic Center, which is also rich in bright discrete gamma-ray sources.
Furthermore intense diffuse gamma-ray emission due to cosmic-ray interactions with interstellar gas and radiation  is detected from the same direction.

A preliminary analysis of the data, taken during the first 11 months of the Fermi satellite operations, is reported.
The diffuse gamma-ray backgrounds and discrete sources, as we know them today, can account for the large majority of the 
detected gamma-ray emission from the Galactic Center. Nevertheless a residual  emission is left, not accounted for by the above models.

An improved model of the Galactic diffuse emission  and a careful evaluation of new  (possibly unresolved) sources (or source populations) will improve the sensitivity
for a DM search.

\end{abstract}

\maketitle

\thispagestyle{fancy}


\section{Indirect Search for Dark Matter with High Energy $\gamma$ rays }
\label{}

After the measurement of the orbital velocities of galaxies within clusters \cite{zwi}, many other
evidences point to the existence of a form of matter which is not coupled to the electromagnetic or to the strong interactions.
Among these it is possible to list:
 the accurate measurements  of the  cosmic microwave background (CMBR) \cite{spe};
 the the galaxies rotation velocities \cite{bor},\cite{per};
 the gravitational lensing \cite{tys};
 the abundances of light elements \cite{oli}; 
and the large scale structures \cite{teg}.
With the measurements of the CMBR  \cite{kom}  the  23\% of the Universe   is estimated to be  Cold  Dark Matter (DM), while only 4\%  \emph{ordinary}  baryonic matter.

DM is coupled with gravitational interaction, and via this force to the ordinary matter. 
As a matter of fact N-body simulations predict that DM is arranged in large structures called \emph{halos} (\cite{spr}).
The DM density in a halo is large at the centre, and decreases with the distance.
The density profile in the central part of the halo  is not experimentally known and is studied by means of fits of  N-body simulations  or 
with  analytic approaches  \cite{lap}.


Non-gravitational DM couplings are studied with:

\begin{itemize}
\item the direct search for DM scattering on ordinary matter;
\item     the indirect study of DM annihilation via the secondary products, both
     charged and neutral (e$^{+}$ , $\bar{p}$, $\bar{d}$, $\nu$ , $\gamma$ rays and lower frequency electromagnetic radiation).
\end{itemize}


DM particles  might produce gamma rays:
\begin{itemize}
\item if the DM particles self-annihilate in pairs, possibly following the scheme in Fig. 1. This yields a continuum gamma-ray emission, which is  produced by hadronization (or final state radiation) of the annihilation products and has a  cut-off at the DM particle mass. The direct production of two gamma rays  as annihilation products is suppressed in many models (10$^{-3}$ - 10$^{-4}$ of the continuum);
\item if pseudo-stable DM particles decay in gammas. In this case the  gamma-ray flux is proportional to the DM particle density. The  DM decay constant is  bound to be larger than 10$^{26}$ sec and is model dependent \cite{ber}.
\end{itemize}

\begin{figure}[h]
\includegraphics[width=8.3cm]{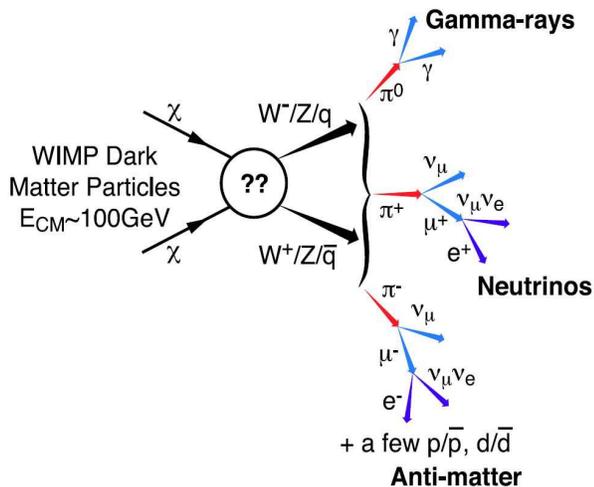}
\caption{gamma-ray production scheme for DM annihilation. Reproduced from \cite{Baltz}.
If the DM particles self-annihilate and produce quarks,
leptons and gauge bosons, then an indirect search is possible by searching for the 
secondary products. For weakly interacting massive particles (WIMPs) of Maiorana type the heavy fermions pairs, such as $b \bar{b}$,$t \bar{t}$ ,$\tau^{+} \tau^{-} $, are favoured as annihilation products. 
The annihilation into two gamma rays, for some models,  is loop-suppressed, with a branching ratio of
10$^{-3}$ -10$^{-4}$. Gamma-ray emission is expected after the heavy fermions hadronize. The 
gamma-ray emission energy spectrum is a continuum with curved shape (in log-log scale) and a sharp cut-off at the DM particle mass.}
\end{figure}

The gamma-ray flux from DM particle annihilation can be decomposed into a particle physics factor and an astrophysical one.
The gamma-ray flux from annihilation can be written as:
$$ \Phi_{WIMP}(E,\psi) \ =  $$
$$ \ = \ \frac{1}{2}  \ \frac{<\sigma v>}{4\pi} \ \frac{1}{m^{2}_{WIMP}}  \sum B_{f}  \frac{dN_{\gamma,f}}{dE_{\gamma}}   \ \int_{los} \rho(l)^{2}dl(\psi)  $$
where $\sigma$ is the annihilation cross-section, $v$ is the relative velocity of the DM particles , $\frac{dN_{\gamma,f}}{dE_{\gamma}}$ is 
the number of gamma rays emitted per annihilation for each annihilation channel f, B$_{f}$ is the branching ratio of the annihilation channel f and
M$_{WIMP}$ is the mass of the DM particle candidate. $\rho$(l) is the DM density profile, ie the density of DM as a function of the distance from the halo center.   $\rho$(l)$^{2}$ is integrated along the observer's line-of-sight ($los$).
The gamma-ray flux from decay can be written as:
$$ \frac{d\phi_{\gamma}}{dE_{\gamma}} \ = \ \frac{\Gamma}{4\pi m_{DM}} \ \frac{dN_{\gamma}}{dE_{\gamma}}  \ \int_{los} \rho(l)dl(\psi) $$
where $\Gamma$ is the decay constant, $\frac{dN_{\gamma}}{dE_{\gamma}}$ is 
the number of gamma rays emitted for each decay  and
M$_{DM}$ is the mass of the DM particle candidate. $\rho_{DM}$ is the DM density profile.

The strong dependence of the emitted $\gamma$-ray flux on the DM density profile is evident in the annihilation case, smaller in the decay one.
In fact in the decay scenario $\rho$, not its square, is integrated along the observer's line-of-sight ($los$, see Fig. 2).

In order to search for a gamma-ray signal from DM annihilation (or decay), 
we observe the region from which we expect the largest signals or the best signal-to-noise ratio,
where the noise is the gamma-ray background.

Our Galaxy could be embedded in a DM halo  with the central density enhancement co-located with the Galactic Center. 
As the  DM gamma-ray emission is expected to be a function of the DM density profile 
the Galactic Center is expected to be the brightest source of gamma rays from DM annihilation (see Fig. 2 and 3) .
It might be possible to detect the relatively large DM gamma emission (\cite{mor}, \cite{ces}, \cite{lp2}),
despite the bright gamma-ray  emission from astrophysical processes, which  is generated both by discrete sources and diffuse backgrounds, in the Galactic Center region

A better ratio between the Dark and baryonic matter, and then a weaker gamma-ray background,
is foreseen for the Dwarf Spheroidal  Galaxies, or in DM sub-structures in the
halo \cite{far}.

\begin{figure}[h]
\includegraphics[width=8.8cm]{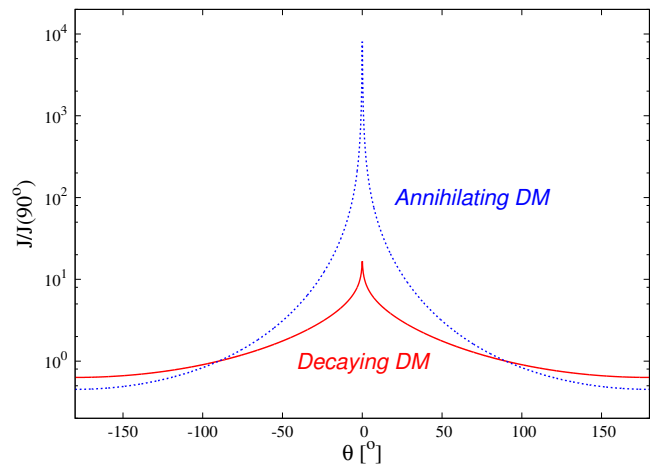}
\caption{Comparison of expected gamma-ray emission profile in the case of annihilating (proportional to the square of the density) and decaying (proportional to the density) Dark. The angular profile of the gamma-ray signal is shown, as function of the angle $\theta$ to the
centre of the galaxy for a Navarro-Frenk-White (NFW) halo distribution for decaying DM, solid (red) line,
compared to the case of self-annihilating DM, dashed (blue) line. Both signals have
been normalised to their values at the Galactic poles, $\theta$ = $\pm$90$^{\circ}$. Reproduced from  \cite{ber}.}
\end{figure}

\begin{figure}[h]
\includegraphics[width=7cm]{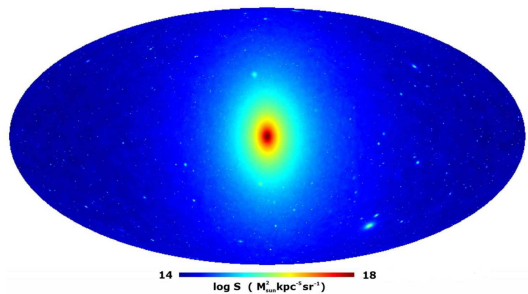}
\caption{Simulated gamma emission from an annihilating  DM halo. The total surface brightness from all components together
 (the main halo + all resolved subhalos + all unresolved subhalos down to the free streaming limit) is shown. 
The surface brightness is in color code. Reproduced from \cite{spr}.}
\end{figure}


\section{Fermi Large Area Telescope}

The Fermi Gamma-ray Space Telescope is a space observatory for the study of gamma-ray emission from astrophysical sources.
It is  equipped  with two powerful instruments:
\begin{enumerate}
\item the Large Area Telescope (LAT), a pair-conversion telescope, composed of  a precision silicon-tracker (18 double-sided layers)  
and a calorimeter (8.6 radiation lengths vertical depth), each of which consisting of a 4 $\times$ 4 array of 16 modules and surrounded by a segmented anti-coincidence detector (ACD).
LAT is operated as a  gamma-ray imager  in the energy range between 20 MeV to over 300 GeV;
\item the gamma-ray Burst Monitor (GBM), a detector covering the 8 keV-40 MeV energy range, devoted to the study of the gamma-ray Bursts.
\end{enumerate}
Detailed descriptions of the Fermi  observatory  can be found  in \cite{Atwood} and the LAT on-orbit calibration is reported in \cite{lat2}.

Studies of the  gamma-ray sources in the energy range between  hundreds of MeV and tens of GeV were performed by means of the SAS 2 and COS B satellites  in the years 1972 - 1982 and
with  the Energetic gamma-ray Experiment Telescope (EGRET) onboard of the Compton gamma-ray Observatory between 1991 and 2000.
EGRET detected 271 sources \cite{Hartman}. An half of the EGRET sources are  unidentified, mainly because of the relatively large errors associated with  the source locations.
The sources detected with EGRET are mainly members of two classes: Active Galactic Nuclei (AGNs) of FSRQ (Flat Spectrum Radio Quasars) and BL Lac types, with powerful relativistic jets of plasma, and
pulsars (spinning neutron stars, with powerful magnetic fields, capable of accelerating particles up to the high energy regime).
EGRET was able to detect both discrete sources and  diffuse  gamma-ray emissions such as the extra-galactic  gamma-ray  background and the very  intense Galactic component.

The Large Area Telescope has an effective area five times larger, a much better angular resolution,
and  a  sensitivity more than 10 times better than its predecessor EGRET.
The Fermi observatory  has several scientific objectives, which span many topics of astrophysics and fundamental physics:
(1) the  detailed study of pulsars, AGNs, diffuse emissions and gamma-ray emission from nearby bodies;
(2) the  study of gamma-ray Bursts, up to   GeV energies with the LAT and in the better studied keV- MeV range by means of the GBM;
(3)  the search for new classes of gamma-ray emitters; 
(4)  the possible signals of new physics. The indirect search for DM particles and the investigation  of their nature are  major research topics for Fermi.
Here we report an update of the indirect search for DM from the Galactic Center (GC).

\section{The Galactic Center  during the first year of Fermi}

Before the launch of the Fermi observatory, the sensitivity of the LAT instrument for the indirect search 
for DM was estimated in \cite{Baltz}.
In this publication many possible targets for the observation of gamma-ray emission from annihilating DM are considered and the 
possible limits on the velocity-averaged annihilation cross-section ($<$$\sigma$v$>$ parameter) are given, as a function of the DM particle mass.
For the Galactic Center, the considered gamma-ray background was the Galactic Diffuse emission, but not discrete bright gamma-ray sources.
The results are obtained by means of $\chi^{2}$ analysis.
The plot in Fig. 4 shows the constraints in $<$$\sigma$v$>$  vs WIMP mass for an assumed pure $b\bar{b}$ annihilation channel.

A likelihood analysis was instead  performed in \cite{vit1},  on simulations of both 
annihilating DM gamma-ray emission and gamma-ray  background (both Galactic diffuse emission and the known discrete gamma-ray sources).
With this model it was possible to conclude that the Fermi/LAT is sensitive to many benchmark realizations of a DM source,
under the assumption that the  simulation of the gamma-ray background is realistic.
For example a DM particle with mass = 50 GeV and a $<$$\sigma$v$>$ = 3$\times$10$^{−26}$cm$^{3}$s$^{-1}$  would have provided $\approx$280 events, 
within a ROI of 1.8$^{\circ}$  radius, in 30 Ms.
In the same time the simulated background would have generated $\approx$20000 events, and a detection at 5$\sigma$ would have been possible, with a maximum likelihood analysis,
which exploits both  the spatial and the spectral  information.
Unfortunately the simulated background did not reproduce the real complexity of the region.

\begin{figure}[h]
\includegraphics[width=8.0cm]{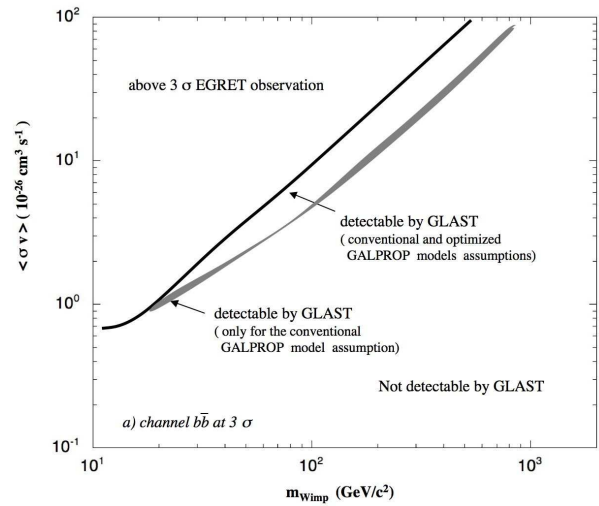}
\caption{ The Fermi gamma-ray Satellite sensitivity for DM
indirect searches was investigated in \cite{Baltz}. For the
GC the DM mass was considered between 10 and
1000 GeV, while the $<$$\sigma$v$>$ parameter between 0.5 and
100$\times$10$^{−26}$cm$^{3}$ s$^{-1}$ . The Galactic diffuse emission (both
conventional and optimized GALPROP models) was assumed as background, while the sources were not included as background, and a $\chi^{2}$ analysis
was applied. In Fig 3 is reported the region of the  $<$$\sigma$v$>$ vs DM mass plane which can be sampled with Fermi/LAT in 5 years, with the observations of the Galactic Center. For this plot a pure $b\bar{b}$ annihilation channel and a NFW spatial distribution are assumed.
}
\end{figure}

The first  release of results on the Galactic Center region obtained with the Fermi/LAT instrument, was made with   the $\emph{Fermi \ Bright  \ Source \ List}$ publication (\cite{bsl}, see Table 1).
The Fermi/LAT in 3 months produced a deep and well-resolved map of the $\gamma$-ray sky, thus also providing much better localization of the previous known gamma-ray sources.
205 sources were detected during the first 3 months with  statistical significance larger than 10 $\sigma$.
The minimum flux for detection of this quality changed from the Galactic plane regions ($\Phi$(E$>$100 MeV)=0.4$\times$10$^{-6}$cm$^{-2}$s$^{-1}$) to the Galactic poles ($\Phi$(E$>$100 MeV)=0.05$\times$10$^{-6}$cm$^{-2}$s$^{-1}$),
as a function of the intensity of the bright astrophysical gamma-ray background (mainly the galactic component). 
In \cite{bsl} the source 0FGL J1746.0-29.00  is the  source closest to the dynamical center of the  of the Galaxy.
0FGL J1746.0-2900 was reported to be marginally variable in \cite{bsl}, but this result was not confirmed with larger statistics.

0FGL J1746.0-2900 is likely associated to 3EG J1746-2851, an unidentified EGRET source.
A re-analysis of the EGRET data \cite{pohl} indicated for this source a steady integral flux
$\Phi$($>$100 MeV)=(118$\pm$73)$\times$10$^{-8}$cm$^{-2}$s$^{-1}$, with a position  displaced from the exact
Galactic center toward positive Galactic longitudes.

The TeV source in the Galactic Center (HESS J1745-290), a bright source above 100 GeV, is notably located near  0FGL J1746.0-2900. 
The energy spectrum measured with H.E.S.S. \cite{aha} over the three years of data taking is compatible
with both a power law spectrum with an exponential cut-off (photon index of 2.10$\pm$0.04$_{stat}\pm $0.10$_{syst}$ and a cut-off energy at 15.7$\pm$3.4$_{stat}$
$\pm$2.5$_{syst}$ TeV) and a broken power law spectrum (spectral indices of 2.02$\pm$0.08$_{stat} \pm$ 0.10$_{syst}$ and 2.63$\pm$0.14$_{stat} \pm$0.10$_{syst}$
with a break energy at 2.57$\pm$0.19$_{stat}\pm$ 0.44$_{syst}$ TeV). No significant flux variation was found. 
The identification of the TeV gamma-ray source is still pending, although the DM annihilation hypothesis for the TeV emission is not favoured \cite{aha2}.
Furthermore the H.E.S.S. collaboration reported also weaker $diffuse$  emission from the central part of the Galactic Ridge \cite{rid}.

\begin{table}[h]
\begin{tabular}{|c|c|c|c|c|}
\hline
Source & l & b  & $\Theta_{95}$ & Int.Flux \\ 
        &  &    &   &(1$<$E$<$100 GeV) \\
0FGL id & ($^\circ$) &  ($^\circ$)  &  ($^\circ$) & (10$^{-8}$cm$^{-2}$s$^{-1}$ ) \\
\hline
 J1732.8-3135 & 356.287 & 0.920  & 0.087 & 3.89$\pm$0.33 \\
 J1741.4-3046 & 357.959 & -0.189  & 0.197 & 2.00$\pm$0.31 \\
 J1746.0-2900 & 359.988 & -0.111  & 0.068 & 7.92$\pm$0.47 \\

\hline
\end{tabular}
\caption{Sources of the Fermi Bright Sources List, with -5$^{\circ}<$l$<$5$^{\circ}$ and  -5$^{\circ}<$b$<$5$^{\circ}$. The source identifier, the galactic longitude and latitude, the   radius of the 95\% source location confidence region and the integral flux between 1 and 100 GeV are reported. 
The Bright Sources were detected with a statistical significance  above 10$\sigma$ in 3months.}
\end{table}

During summer 2009 a first preliminary analysis of the Galactic Center data, for the indirect search for DM, was reported both at the TeVPA \cite{meu} and ICRC 2009 \cite{vit2} international conferences.
For those results  the likelihood analysis was used:
\begin{itemize} 
\item a Region-Of-Interest (ROI) of 1$^{\circ} \times $1$^{\circ}$ was considered;
\item the ROI was centred at the position RA = 266.46$^{\circ}$, Dec=-28.97$^{\circ}$;
\item the Data taken during the first 8 months (8/2008-4/2009) have been used;
\item the  events were selected to have energy  between 200 MeV and 40 GeV;
\item only events classified of "diffuse" class were selected for the analysis.
\end{itemize}
For that 8 months analysis data were binned into a single spatial bin, then no spatial information was considered.
Likelihood analysis for the LAT data requires a  model of the background (both diffuse emission and known sources).
At that time  a broken power law was assumed as the model of the  gamma-ray emission from this small ROI, as it was well-describing the observed data.
The broken  power-law spectrum was found to have  integral flux
 between 100 MeV and 100 GeV  $\Phi(0.1-100 GeV)$ = (1.22$\pm$0.02) 10$^{-6}$ cm$^{-2}$s$^{-1}$, first spectral index $\Gamma_{1}$= -1.38 $\pm$ 0.04,
second spectral index  $\Gamma_{1}$= -2.60$\pm$ 0.05, energy break at E$_{break}$ = (1.6$\pm$0.1)GeV.
Then a component  representing the gamma-ray emission of annihilating DM was added to  the likelihood analysis model.
A benchmark DM realization with  mass = 50 GeV  and  a pure $b\bar{b}$ annihilation channel spectrum was considered.
 
With this model it was possible to put an upper limit of  2.43$\pm$0.02 10$^{-7}$ cm$^{-2}$s$^{-1}$ (95\% confidence level, above 100 MeV)  to the integral gamma-ray flux from a
 DM annihilation source, which was assumed to have a Navarro-Frenk-White density profile, and the above reported benchmark mass of 50 GeV.
This gamma-ray flux upper limit implies a limit of 39.8 $\times$  10$^{-26}$ cm$^{3}$s$^{-1}$ to the  $<$$\sigma$v$>$ parameter, if a pure $b\bar{b}$ annihilation channel is assumed.
Nevertheless the analysis is to be  considered  very conservative, because no spatial information was used.

\begin{figure}[h]
\includegraphics[width=8.3cm,height=5.cm]{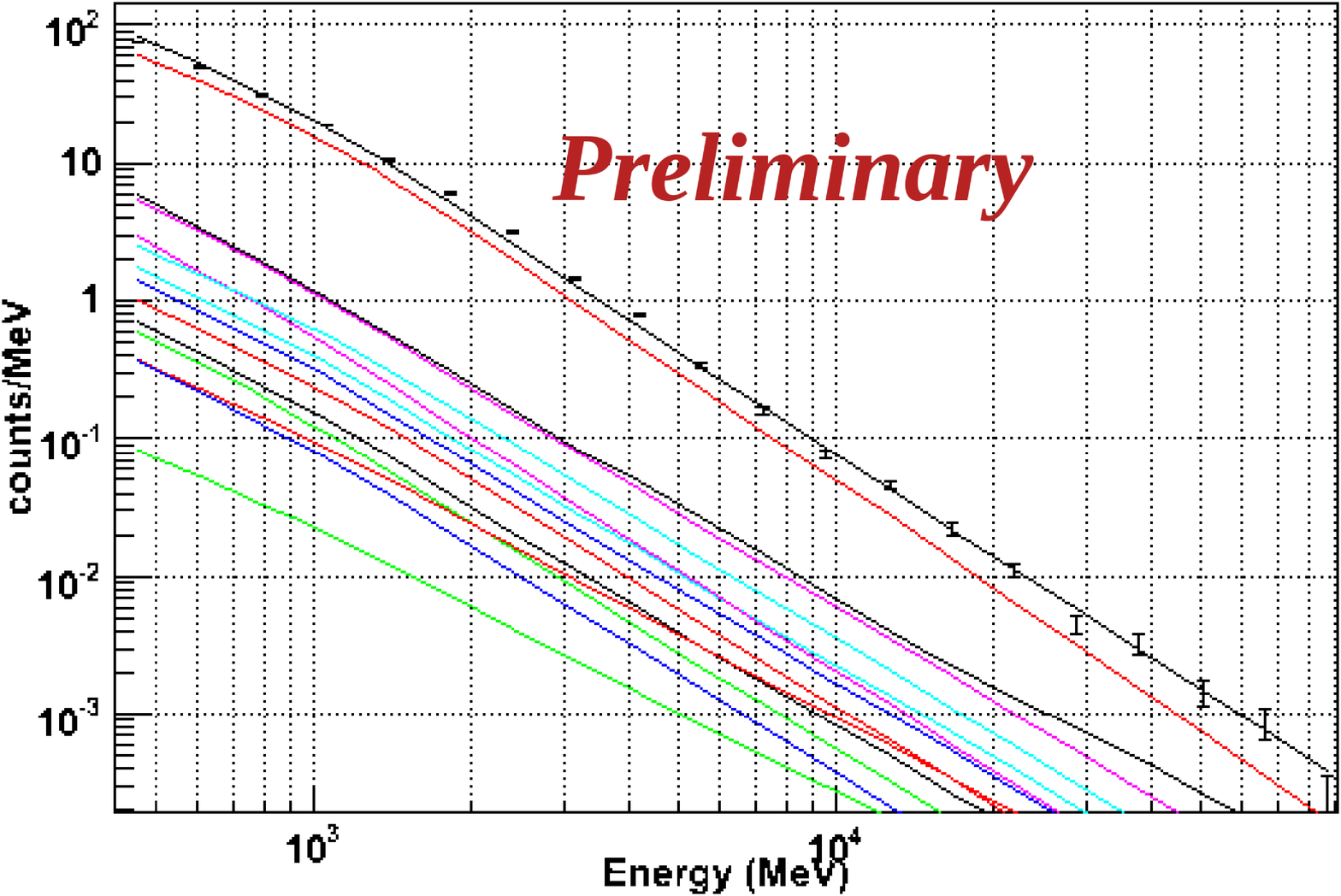}
\caption{Spectra from the likelihood analysis of the Fermi/LAT data (number of counts vs reconstructed energy).
The likelihood analysis is the standard one used with the LAT data.
The main analysis steps are:
(1) to select data of high quality (selection cuts on events energy, zenith angle, reconstruction and classification quality); (2) to build a emission model of the region, based on the previous knowledge and experimental evidence of new excesses with enough statistical significance; (3) to apply the likelihood analysis to the data and the considered model. For each model component a fit of the free parameters and the computation  the statistical significance is obtained.
Here in the plot, from highest to the lowest:
the black point are the observed data;
the black line is the sum of all the components;
the red line  is the Galactic diffuse emission;  the lower black line is the isotropic extragalactic;
the other components are the sources detected. These results are preliminary.}
\end{figure}

\begin{figure}[h]
\includegraphics[width=8.3cm,height=4.2cm]{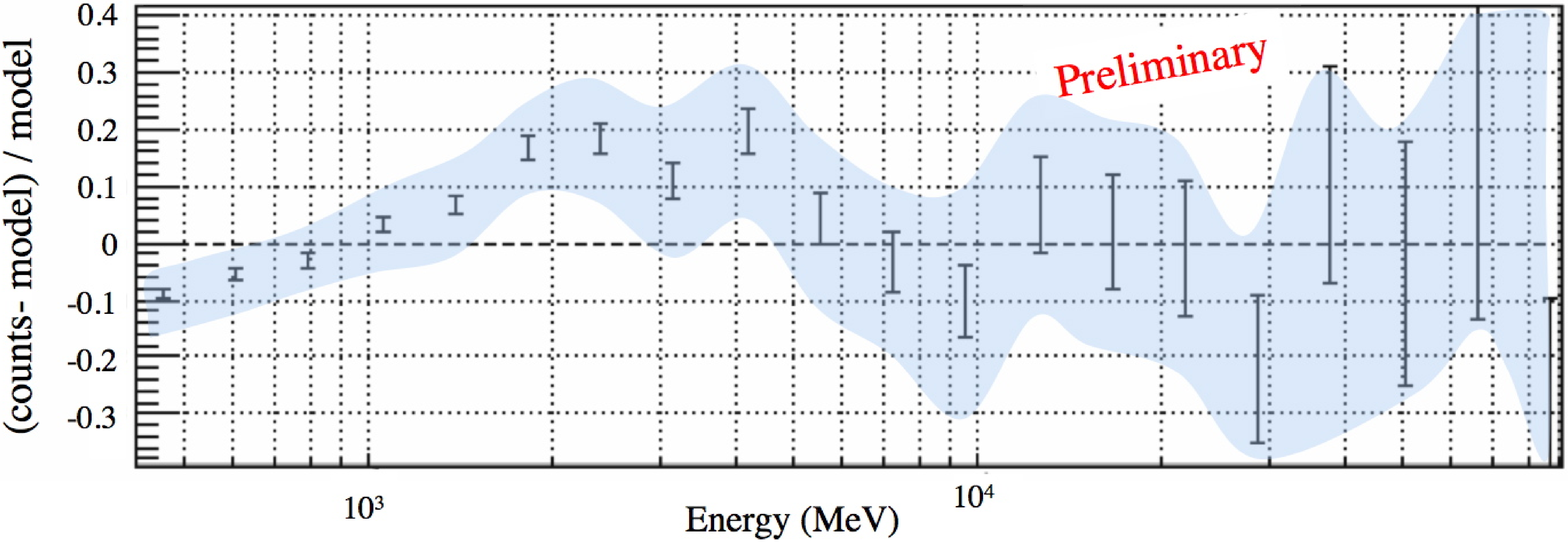}
\caption{Residuals ( (exp.data - model)/model ) of the above likelihood analysis. 
For this residual spectrum the model and  data were integrated over the 7$^{\circ} \times$7$^{\circ}$  ROI.
The  residual spectrum suggests that an unmodeled excess is  present in the  $\approx$2-5  GeV range.
The blue area  shows the systematic errors on the effective area. These results are preliminary.}
\end{figure}

\section{Preliminary analysis of first 11 months data}

The reported results were obtained with a binned  likelihood analysis, performed by means of the tools developed by the Fermi/LAT collaboration (gtlike, from the Fermi analysis tools \cite{tools}).  
For this  analysis:
\begin{itemize} 
\item a ROI of 7$^{\circ} \times $7$^{\circ}$ was considered. This ROI was used in order to minimize the background contribution 
and to avoid significant leakage of the gamma-ray signal  under study;
\item the ROI was centred at the position RA = 266.46$^{\circ}$, Dec=-28.97$^{\circ}$, ie the position of the brightest source;
\item the Data taken during the first 11 months (8/2008-7/2009) have been used;
\item the  events were selected to have energy  between 400 MeV and 100 GeV;
\item only events classified of \emph{diffuse} class and  which converted in the \emph{front} part of the tracker have been selected for the analysis. The selection 
in energy, event classification and conversion provided us with events with very well reconstructed incoming direction and data have been binned into a 100$\times$100 bins map;
\item the IRF and the events classification are those relative to the Pass6V3 version of the Fermi/LAT analysis software.
\end{itemize}

In order to perform the likelihood analysis for the LAT data, a  model of the already known sources and the diffuse  background should be built.
The model in use for the presented analysis contains  11 sources in the Fermi  1 year catalog (to be published) which are located  into  the considered ROI, or located  very close to the ROI boundaries and have a significant fraction of their flux leaking within the studied region.
These sources have a point-like spatial model and a spectrum in the form of a power-law.
The model also contain the diffuse gamma-ray background which is composed by two components:
\begin{enumerate}
\item the Galactic Diffuse gamma-ray background. The observed  Galactic Diffuse emission was modelled  by means of the GALPROP code \cite{str} and \cite{str2}, and 
the  realization of the galactic emission named gll$_{-}$iem$_{-}$54$_{-}$87Xexph7S.fit was used. During 
the likelihood maximization only the normalization of the  GALPROP model is varied, not its components;
\item the Isotropic Background. This component should account for both the Extragalactic gamma-ray emission and  residual charged particles. It is modelled as an isotropic emission with a
template spectrum.
\end{enumerate}

The obtained results are illustrated in Fig. 5 and 6.
The bulk of the gamma-ray emission from this region is explained by means of the above described components, but a residual emission is left.
The systematic uncertainty of the effective area of the LAT is ~10\%   at 100 MeV, decreasing to 5\% at 560 MeV and increasing to 20\% at 10  GeV. 
 This uncertainty propagates to the model predictions and should  be considered in interpreting the residual spectrum in Fig 6.

\section {Conclusions}

We have reported a preliminary analysis of the Fermi/LAT observations of the Galactic Center.
This analysis was focused on the   indirect search for DM.
From this preliminary analysis the following conclusion are obtained:
\begin{itemize}
\item Any attempt to disentangle a potential DM signal from the galactic center region requires deep understanding of the conventional astrophysics background;
\item The bulk of the gamma-ray emission from the GC region is explained with the detected  sources and the Galactic Diffuse emission model, but ...
\item ... a residual gamma-ray emission is left, not accounted for by the above models.
\end{itemize}

Improved modelling of the Galactic 
diffuse model as well as the potential contribution from other 
astrophysical sources (for instance unresolved point sources) could 
provide a better description of the data. Analyses are under-way to 
investigate these possibilities.

\bigskip 
\begin{acknowledgments}

  The $Fermi$ LAT Collaboration acknowledges the generous support of a 
number of agencies and institutes that have supported the $Fermi$ LAT 
Collaboration. These include the National Aeronautics and Space 
Administration and the Department of Energy in the United States, the 
Commissariat \`{a} l'Energie Atomique and the Centre National de la 
Recherche Scientifique / Institut National de Physique Nucl\'{e}aire et de 
Physique des Particules in France, the Agenzia Spaziale Italiana and the 
Istituto Nazionale di Fisica Nucleare in Italy, the Ministry of 
Education, Culture, Sports, Science and Technology (MEXT), High Energy 
Accelerator Research Organization (KEK) and Japan Aerospace Exploration 
Agency (JAXA) in Japan, and the K.\ A.\ Wallenberg Foundation, the 
Swedish Research Council and the Swedish National Space Board in Sweden. 
JC is Royal Swedish Academy of Sciences Research fellow supported by a 
grant from the K.\ A.\ Wallenberg foundation.

\end{acknowledgments}

\bigskip 

\end{document}